\documentclass{article}
\usepackage{booktabs}
\usepackage{tabularx}
\usepackage{array, amssymb}
\usepackage{spconf,amsmath,graphicx,hyperref}
\makeatletter
\newcommand{\tightdisplayskips}{%
  \setlength{\abovedisplayskip}{5pt plus 2pt minus 2pt}%
  \setlength{\belowdisplayskip}{5pt plus 2pt minus 2pt}%
  \setlength{\abovedisplayshortskip}{2pt plus 2pt minus 1pt}%
  \setlength{\belowdisplayshortskip}{2pt plus 2pt minus 1pt}}
\g@addto@macro\normalsize{\tightdisplayskips}
\g@addto@macro\small{\tightdisplayskips}
\makeatother
\title{Personalising Parametric Audio Effect Recommendation for Vocals from Parameter-only Style Priors}
\name{Soumya Sai Vanka$^{\star}$ \quad Hayato Yamakawa$^{\dagger}$ \quad Kazunobu Kondo$^{\dagger}$ \quad Yu Takahashi$^{\dagger}$ \quad György Fazekas$^{\star}$}
\address{$^{\star}$ C4DM, Queen Mary University of London UK, $^{\dagger}$RDD, Yamaha Corporation, Japan}
\begin{document}
\maketitle
\begin{abstract}
Audio effect recommendation has mostly modelled style through reference audio and audio-domain losses. Because the parameter-to-audio map of parametric effects is many-to-one, such systems can match a reference without reflecting how a given engineer works. We instead model an engineer's style as a distribution over effect-chain parameters, which matches the data engineers can realistically share: console recalls store parameters without the audio they were applied to. Under a label-shift assumption, a personalised recommendation follows in closed form from a generic content model, a parameter population model and an engineer prior, with no paired engineer audio and no tuned weights. For vocal chains, the content signal is small but real, and concentrated in high-frequency EQ, dynamics and reverb. A forced-choice test with $14$ listeners found that the style-prior and content-only recommendations were distinguishable ($89.3\%$), with results consistent with prior width affecting the balance between style and content, while parameter-space cosine does not predict perceived difference.

\end{abstract}
\section{Introduction}
\label{sec:intro}
Recent work on mixing, mastering and audio effect recommendation models style through reference audio~\cite{ycy2025ito, koo2023music, sai2024diff, steinmetz2024st, steinmetz2022style}, scored with audio-domain losses such as MRSTFT or MIR descriptors~\cite{ycy2025ito, sai2024diff}. For parametric systems~\cite{ycy2025ito, sai2024diff, steinmetz2024st}, the parameter-to-audio map is many-to-one~\cite{hayes2025audio}, so a system optimised on audio alone can match the reference without reflecting how a given engineer works. We instead define style as an engineer's way of using the effect chain: which processors they engage and the values they habitually choose. A recommendation is an editable starting point, so its usefulness depends on how closely its parameters match the engineer's practice as well as on the audio it produces.\\
The realistic data for this is asymmetric. Style is usually stored as console recalls, per-scene snapshots of effect-chain parameters, without the audio: session audio is often client property and rarely shared, while consoles, DAWs and plugins keep parameters independently of it. An internal audit of Yamaha console recalls (\textbf{7} engineers, \textbf{200+} scenes) shows consistent per-engineer patterns in effect engagement, EQ direction and magnitude, and compression ratio and threshold. A prior estimated from recalls therefore captures an engineer's tendencies but not how settings should change with the input, while a generic content model learned from paired audio and parameters captures content-dependence but not identity. We combine the two under a label-shift formulation that gives a closed-form personalised recommendation with no paired engineer audio and no tuned weights, and evaluate it for vocal effects in parameter space and perception using synthetic style priors.
\subsection{Problem Formulation}
Let $\theta \in \Theta \subseteq \mathbb{R}^d$ be the parameters of a fixed vocal chain (EQ, dynamics, delay, reverb) in a bounded canonical space, and $a$ the features of the dry vocal. For engineer $e$ we seek $p_e(\theta \mid a)$. The data provide two estimable objects but never their combination: recalls give style priors $p_e(\theta)$ without audio, and a paired public corpus gives a content model $p(\theta \mid a)$ and its marginal $p(\theta)$ without engineer identity.\\
\textbf{Label-Shift Assumption}: Since recalls contain no audio, the engineer-specific coupling $p_e(a \mid \theta)$ is unidentifiable, so we assume it is shared,
{\small
\begin{equation}
    p_e(a \mid \theta) = p(a \mid \theta) \qquad \text{for all } e,
    \label{eq:label_shift}
\end{equation}}%
a label-shift assumption with $\theta$ as label and $a$ as covariate~\cite{saerens2002adjusting,lipton2018detecting}. It constrains the coupling $a \mid \theta$, not the decision $\theta \mid a$, so engineers can still favour different configurations, $p_e(\theta) \neq p(\theta)$. It fails if an engineer's material differs systematically from the corpus, e.g.\ in genre or voice type.\\
\textbf{Combination Rule}: Bayes' rule within engineer $e$, Eq.~\eqref{eq:label_shift} and $p(a\mid\theta) = p(\theta\mid a)p(a)/p(\theta)$ give
{\small
\begin{equation}
    p_e(\theta \mid a)
    = \frac{p(a \mid \theta)\,p_e(\theta)}{p_e(a)}
    \propto \frac{p(\theta \mid a)}{p(\theta)}\,p_e(\theta),
    \label{eq:combination_rule}
\end{equation}}%
with $p(a)/p_e(a)$ absorbed into normalisation. The ratio $p(\theta\mid a)/p(\theta)$ is the content update: dividing by $p(\theta)$ removes the generic preference already in $p(\theta \mid a)$, so it is not double-counted when re-anchoring onto $p_e(\theta)$. Setting $p_e(\theta) = p(\theta)$ recovers the generic recommendation $p(\theta \mid a)$.\\
\textbf{Gaussian Instantiation}: With $p(\theta \mid a) = \mathcal{N}(m(a), S)$, $p(\theta) = \mathcal{N}(\mu_\theta, \Sigma_{\theta\theta})$ and $p_e(\theta) = \mathcal{N}(\mu_e, \Sigma_e)$, Eq.~\eqref{eq:combination_rule} is Gaussian with precision $\Lambda$ and mean $\mu$,
{\small
\begin{align}
    \Lambda &= S^{-1} - \Sigma_{\theta\theta}^{-1} + \Sigma_e^{-1},
    \label{eq:gaussian_precision}\\
    \Lambda\mu &= S^{-1}m(a) - \Sigma_{\theta\theta}^{-1}\mu_\theta + \Sigma_e^{-1}\mu_e.
    \label{eq:gaussian_mean}
\end{align}}%
This is a precision-weighted estimate with no tuned interpolation weight: it follows the engineer prior where $\Sigma_e$ is small and the audio where $S$ is small. Supplementary material can be found \href{https://sai-soum.github.io/projects/vocal-fx-personalisation.html}{here}.
\begin{figure}[t]
    \centering
    \includegraphics[width=0.8\columnwidth]{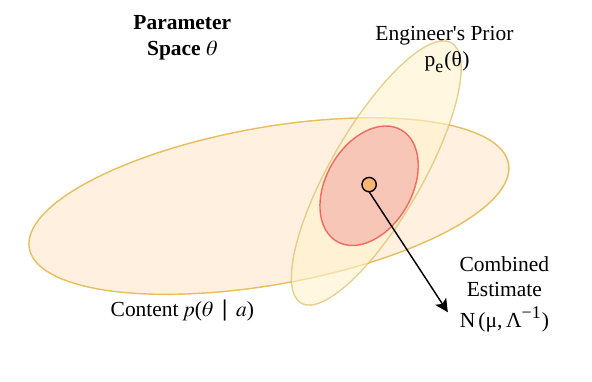}
    \caption{Gaussian case of Eq.~\eqref{eq:combination_rule}: the content model and engineer prior combine, with the population base rate divided out.}
    \label{fig:gaussian_combination}
\end{figure}
\section{EXPERIMENTAL SETUP}
\textbf{Parameter ($\theta$) Space}: We use the 70 filtered dry-wet vocal track pairs from 52 independent MedleyDB~\cite{bittner2014medleydb, bittner2016medleydb} songs and the corresponding 70 vocal effect chains derived in DiffVox~\cite{ycy2025diffvox} as our dataset for the content and population models. The 70 track-level parameter vectors are used for population estimation, while the song identity defines the independent unit for validation. DiffVox integrates parametric EQ, a compressor-expander, a send-based delay and an FDN reverb via differentiable implementations fitted to each dry-wet pair by gradient-based optimisation. The chain's 151 parameters are masked to 130 non-redundant dimensions (21 redundant lower-triangular entries of the FDN's mixing matrix are dropped); full processor and parameter definitions are given in the supplementary material. All modelling is done in this native 130-dim coordinate, never in physical units (dB, Hz, ms) directly, since several parameters are bounded and a raw/logit coordinate is closer to unconstrained-Gaussian than the physical one. \\
\textbf{Content Model $p(\theta \mid a)$}: The content model predicts 130 parameters from audio using only 70 paired examples from 52 independent songs, so we keep its capacity small at every step. \textit{Features.} We compute three feature sets on the loudness-normalised dry vocal, over 12~s windows with 5~s overlap: 37 hand-designed statistical descriptors, the 128-dimensional Fx-Encoder++ embedding~\cite{yeh2025fx}, and the 1024-dimensional AfxRep embedding~\cite{steinmetz2024st}, each summarised by its mean and standard deviation since dynamics respond to variability over time (full descriptor list in the supplementary material), then reduced to six PCA components (three on the means, three on the standard deviations) for an 18-dimensional combined representation. \textit{Regression.} We fit PLS regression and choose its rank from 1 to 6 by held-out $R^2$ under leave-one-song-out cross-validation, grouped by song since some MedleyDB songs contribute several stems sharing one mix. Throughout, $\theta_i$ denotes the DiffVox-fitted parameters of track $i$, treated as ground truth, and $\hat{\theta}_i$ the held-out prediction; for an actual recommendation the model is refit on all 70 tracks. \textit{Residual covariance.} We estimate $S$ from the held-out residuals $\theta_i - \hat{\theta}_i$, not the training fit, which would overstate the content model's confidence, and apply Ledoit-Wolf shrinkage~\cite{ledoit2004well}. We require $S$ and $\Sigma_{\theta\theta}$ to be positive-definite and enforce $S \preceq \Sigma_{\theta\theta}$, since audio cannot make us less certain than the population alone. Since inversion reverses Loewner ordering, $S^{-1} - \Sigma_{\theta\theta}^{-1} \succeq 0$; with $\Sigma_e \succ 0$, this gives $\Lambda \succ 0$ in Eq.~\eqref{eq:gaussian_precision}. \\
\textbf{Population Model $p(\theta)$}: With 70 DiffVox-fitted vocal chains in 130 dimensions the sample covariance is singular, so we estimate $\mu_\theta$ and $\Sigma_{\theta\theta}$ by Ledoit-Wolf shrinkage. In a leave-one-song-out comparison, each covariance estimator was fitted on the remaining songs and its Gaussian log-likelihood was evaluated separately on the presets from the held-out song. Ledoit-Wolf outperformed PCA truncation for 58 of the 70 held-out presets and keeps all 130 dimensions, needed because Eq.~\eqref{eq:gaussian_precision} inverts $\Sigma_{\theta\theta}$. We fit the content model blockwise, five separate PLS models over EQ, compressor, delay, FDN and pan/send, each with its own rank, rather than one joint model over all 130 dimensions, since blockwise gives the better content model. Blockwise leaves zeros between blocks in $S$ (unmodelled, not uncorrelated), so a multi-block profile is combined one block at a time.\\
\textbf{Synthetic Style Priors $p_e(\theta)$}: We build synthetic style priors grounded in the directions observed in the real recalls and use them as controlled experimental probes, without being fitted to them, because the underlying DSP differs between the real recalls and the DiffVox presets. The priors constrain only selected EQ and dynamic parameters, so we do not claim to model engineer-specific preferences for delay or reverb. We construct 13 profiles in the same raw coordinate as the model, spanning spectral, dynamic and cross-block targets, plus three vague versions that reuse a confident profile's $\mu_e$ with $\Sigma_e$ widened $25\times$ so that each confident/vague pair isolates the effect of $\Sigma_e$ with $\mu_e$ fixed; each sets a target in physical units on 4 to 10 parameters, with $\Sigma_e$ diagonal. Table~\ref{tab:style_profiles} shows three representative profiles, one per category, used below to illustrate the combination and calibration results; the full set of 13 is in the supplementary material.
\begin{table}
\caption{Three representative style profiles (of the 13 constructed; full list, including the \textit{vague} variants that reuse these three $\mu_e$ with $\Sigma_e$ widened $25\times$, in the supplementary material). EQ gains are low shelf / peak 1 / peak 2 / high shelf.}
\label{tab:style_profiles}
\centering
\footnotesize
\setlength{\tabcolsep}{2.7pt}
\renewcommand{\arraystretch}{0.8}
\begin{tabularx}{\linewidth}{@{}l>{\raggedright\arraybackslash}X@{}}
\hline
\textbf{Profile} & \textbf{Target, in physical units} \\
\hline
carver \textit{(spectral)} & $-4.5/-4.5/-3.5/-3.0$ dB, peak Q 1.6 \\
heavy\_leveller \textit{(dynamic)} & $-30$ dB, $6{:}1$, 5/250 ms, $+4$ dB \\
surgical\_carver \textit{(combination)} & $-4.0/-4.0/-3.0/-2.5$ dB, Q 3.2;
$-22$ dB, $3.5{:}1$, 15/150 ms \\
\hline
\end{tabularx}
\end{table}
\section{RESULTS}
\begin{figure}[t]
    \centering
        \includegraphics[width=\columnwidth]{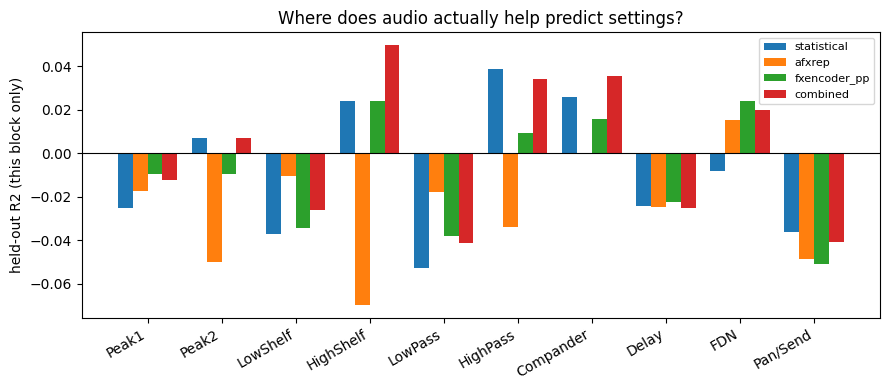}
    \caption{Content model effect block sensitivity by audio features.}
    \label{fig:content_model_blocks}
\end{figure}

\begin{table*}[th]
\centering
\caption{Listening-test procedure and pooled results ($n=14$). Chance is 0.5 for all tests; 62.5\% is needed for $p<0.05$ at 56 observations per condition (64.3\% at T1's 42); $p$ is an exact one-sided binomial test. The full per-condition breakdown (individual profiles and pairs) is in the supplementary material.}
\label{tab:forced_choice}
\footnotesize
\setlength{\tabcolsep}{3pt}
\renewcommand{\arraystretch}{0.9}
\begin{tabularx}{\linewidth}{@{}l>{\raggedright\arraybackslash}X>{\raggedright\arraybackslash}Xccc@{}}
\hline
\textbf{Test} & \textbf{Compares} & \textbf{Establishes} & \textbf{correct} & \textbf{\%} & $\boldsymbol{p}$ \\
\hline
T0 & content-only vs.\ combined rec., ABX vs.\ a repeat of one & combined and content-only recommendations were distinguishable in this test. & 200/224 & 89.3 & $<10^{-4}$ \\
T1 & ref.\ profile $P$ vs.\ probes $P,Q$ (AXY, across songs) & distance predicts perceived difference? & 235/420 & 56.0 & .008 \\
T2 & confident profile vs.\ its vague counterpart, vs.\ content-only & confident and vague priors were distinguishable, consistent with $\Sigma_e$ width affecting displacement. & 101/168 & 60.1 & .005 \\
T3 & a $\chi^2$-flagged combination vs.\ an accepted one & flagged combinations were selected more often as having perceptual problems & 110/168 & 65.5 & $<10^{-4}$ \\
\hline
catch trials & \multicolumn{2}{l}{obviously degraded anchor; attention check, not a result} & 56/56 & 100.0 & $<10^{-4}$ \\
\hline
\end{tabularx}
\end{table*}
\textbf{The Content Signal is Weak but Detectable}: We report held-out $R^2$ pooled over all 130 dimensions rather than averaged per dimension,
$R^2 = 1 - \sum_i \lVert \theta_i - \hat{\theta}_i \rVert^2 / \sum_i \lVert \theta_i - \bar{\theta} \rVert^2$, 
with $\bar{\theta}$ the per-dimension mean of $\theta$ across the 70 tracks. Because $\hat{\theta}_i$ is out-of-sample while $\bar{\theta}$ is not, $R^2$ can be negative, and since the rank is chosen on the same score, $R^2$ compares recipes rather than giving an unbiased absolute fit. To test whether the fit is real we use a permutation gate: shuffle which song's audio goes with which song's parameters, refit 200 times, and see where the real $R^2$ falls, maximising the null over ranks too so the comparison is matched.\\
The signal is small but real and exceeds the matched permutation null. Held-out $R^2$ is $+0.016$ for both Fx-Encoder++ and combined, $+0.006$ for AfxRep and $-0.008$ for the hand-designed statistical features, rising to $+0.019$ for combined under blockwise, with RMSE improving on the $\bar{\theta}$ baseline by at most 0.80\%. All four feature sets outperformed all 200 shuffled nulls, whose best draw reaches only $-0.029$ to $-0.035$, so the effect is not noise.\\
Fig.~\ref{fig:content_model_blocks} shows where the signal sits, per processor stage. Reading the combined feature set, the positive stages are the high shelf ($+0.050$), the compressor ($+0.035$), the high-pass filter ($+0.034$) and the FDN reverb ($+0.020$), with peak 2 marginal ($+0.007$); the low shelf, low-pass, peak 1, delay and pan/send are all negative. Pooling the six EQ stages into a single target hides this, because the negative low- and mid-band stages cancel the positive high-frequency ones. What content predicts is therefore high-frequency shaping, dynamics and reverb, and not the low- or mid-band EQ decisions.\\
An ablation favours mean+std features, the blockwise structure and six PCA components over the alternatives we tried (full comparison in the supplementary material); the main remaining limit is rank, since a rank-$r$ PLS fit predicts within only an $r$-dimensional subspace and $m(a)$ varies across tracks by just 1.4\% to 4.6\% of $\theta$'s real spread on the median dimension.\\
\textbf{The Combination Shows the Expected Parameter-Space Behaviour}:\label{sec:combination} Pull measures how far the result sits between the two inputs, over the engaged dimensions $D'$ where they differ, 
{\small
\begin{equation}
\mathrm{pull}
=
\frac{1}{|D'|}
\sum_{k \in D'}
\frac{
|\mu_{\mathrm{combined},k} - \mu_{e,k}|
}{
|\mu_{e,k} - m(a)_k|
},
\label{eq:pull}
\end{equation}}%
so 0 means it landed on the engineer's target and 1 on the content prediction. Directionality is the fraction of engaged dimensions where $\mu_{\mathrm{combined}}$ moves away from $\mu_\theta$ in the same direction as $\mu_e$, a recommendation is flagged if its squared Mahalanobis distance from $\mu_\theta$, over the profile's own dimensions, exceeds the $\chi^2_{|D|}(0.95)$ threshold under the Gaussian approximation, and separability between two profiles is the cosine between their offsets $\mu_e - \mu_\theta$ on shared dimensions ($+1$ redundant, $-1$ opposed). Separately, setting $\Sigma_e(\kappa) = \kappa \cdot \operatorname{diag}(\Sigma_{\theta\theta})|_D$ and sweeping $\kappa$ expresses a profile's consistency as a fraction of the population's own variance; the full per-profile sweep is given in the supplementary material.\\
The parameter-space Gaussian combination behaves as expected. Widening $\Sigma_e$ moves the result from the style target toward the content prediction: pull rises from 1.0\% to 22.0\% for \texttt{carver}, 2.5\% to 51.2\% for \texttt{surgical\_carver} and 2.9\% to 63.9\% for \texttt{heavy\_\allowbreak leveller}. Directionality averages 97.5\%, so the content model pulls the recommendation back without reversing the engineer's intent, and joint and blockwise agree here. The result sits 3.49 raw units from content-only on average, so personalisation is not swamped by content. The in-distribution check flags 20.5\% of recommendations, always the same three profiles whichever feature set or structure is used, so it is the profile target that is unusual rather than the content model behind it.\\
The $\kappa$ sweep exposes the main weakness of our synthetic priors: confident profiles sit at $\kappa \approx 0.04$, under 5\% of the population's own variance and closer to a fixed point than a believable human range, while vague variants sit near $\kappa \approx 1$. A 75\% pull toward content needs $\kappa$ between 0.69 and 5.15, so with a content model this weak, a style prior must be nearly as wide as the population itself before the audio competes with the engineer's own direction. Sensitivity varies about $5\times$ across profiles and is not predicted by which effect blocks a profile touches; the full sweep and this block-independence result are given in the supplementary material.\\
\textbf{Perceptual Consequences are Only Partly Captured by Parameter Distance}: \label{sec:perceptual} Every result above is computed on the parameter vector $\theta$, not on audio, so a recommendation could be correct in parameter space and inaudible. We ran a listening test with ten lead-vocal stems from the Mixing Secrets library\footnote{\label{cmt}\url{https://cambridge-mt.com/}}, one per song, each rendered under 14 conditions (the content-only prediction $m(a)$ and 13 style profiles) as 5~s excerpts, loudness-matched to EBU R128 at $-24.4$~LUFS with a $-1$~dBTP ceiling to remove level as a cue. Fifteen sessions were submitted. One was excluded because 69\% of its scored trials were answered before any stimulus had played to the end, leaving fourteen analysed listeners, all of whom completed the test on headphones. Each participant answered 74 trials across four tests, taking a median of 37 minutes. Every test is two-alternative forced choice with no ``cannot tell'' option, so chance is exactly 0.5 and 62.5\% is needed for $p<0.05$ at the 56 observations each condition receives. Table~\ref{tab:forced_choice} summarises what each test compares and establishes, with pooled results; the full per-condition breakdown is in the supplementary material. \\
All fourteen analysed listeners passed the level check and all four catch trials. T0 is clear at 89.3\% (weakest individual profile 82.1\%, all four significant on their own), showing that the combined and content-only recommendations were distinguishable in this test. T2 exceeds chance at 60.1\% pooled (significant for two of three pairs), with results consistent with a narrower $\Sigma_e$  producing a larger displacement from the content-only reference as in Sec.~\ref{sec:combination}. T3 exceeds chance at 65.5\% pooled and is significant for all three profiles, showing that flagged combinations were selected more often as having perceptual problems than accepted combinations in this test. T1 is above chance at 56.0\% ($p=0.008$), so listeners are not guessing, but accuracy does not order by parameter-space cosine: the correlation is $\rho = +0.017$ ($p=0.97$). What separates the pairs instead is how much spectral difference survived loudness matching: the two pairs left with under 0.3~dB residual sit at chance (51.2\%, $p=0.46$), while the eight pairs left with 0.56~dB or more are reliably above it (57.1\%, $p=0.005$). Cosine weights all 130 dimensions equally, which perceptually they are not; this limits the separability analysis rather than the recommendation itself. Two pairs sit persistently, though not significantly, below chance; the per-condition numbers and the reference-direction imbalance behind those cells are in the supplementary material.\\
Two caveats bound these conclusions: loudness matching removes a mean of 3.9~LU across the T1 pairs, so effect sizes are lower bounds on the full chain (accuracy does not track the removed loudness, $\rho=-0.083$); and at $n=14$, a true per-condition effect between 50\% and 62.5\% would go undetected.
\section{DISCUSSION AND LIMITATIONS}
Our method supports cheap personalisation on mixing consoles from parameter-only priors: it needs no retraining as new prior data arrives and falls back on the content model alone when prior data is scarce, still recommending reasonable vocal chains.\\
$\kappa$ is an interpretable control, not a blend weight: $\kappa \approx 0.04$ says an engineer always does the same thing and $\kappa \approx 1$ says they vary as much as the population does. The prior need not represent an engineer identity either: $\mu_e$ can instead encode a semantic target while $\Sigma_e$ controls how strongly that target competes with the audio. What generalises beyond this instantiation is the data asymmetry itself, parameters stored without the audio they were applied to, which is the normal condition wherever settings are recalled, so the same factorisation applies to any instrument or processor chain with a paired public corpus to supply the content model.\\
\textit{Limitations.} The style priors are synthetic: thirteen hand-picked archetypes whose $\Sigma_e$ our own sweep shows is implausibly tight ($\kappa \approx 0.04$), and our console audit found only one to three vocal channels for some engineers, so we verify a mechanism and demonstrate proof-of-concept rather than modelling any individual's style. The content model is also weak in absolute terms ($R^2 = +0.019$, 0.80\% better than predicting $\bar{\theta}$); the permutation test establishes the signal is real, but everything downstream inherits its size, and 52 independent songs, not feature choice or model family, is the binding constraint. We inherit DiffVox's fit quality, and $\theta$ remains a many-to-one proxy for audio, this paper's own premise applied to itself.\\
The combination rule itself, Eq.~\eqref{eq:combination_rule}, is exact for arbitrary densities; only its Gaussian instantiation and the resulting diagnostics are approximate, and measurably so: Shapiro--Wilk rejects normality on 105 of 130 dimensions, and a leave-one-out Mahalanobis test rejects $\chi^2_{130}$ (KS$=0.574$, $p=2\times10^{-22}$). We use the Gaussian form for its closed-form, lightweight fit to online personalisation with limited data, not as a claim that parameters are Gaussian; T3 found a perceptual difference between flagged and accepted combinations in this limited test, despite the Gaussianity violation; it does not validate the $\chi^2$ calibration generally. Two smaller gaps remain: an unresolved tie between our two best feature sets, and no range-check on Eq.~\eqref{eq:gaussian_mean}.\\
\textit{Future work.} Since cosine in $\Theta$ does not predict perceived difference (Sec.~\ref{sec:perceptual}), selecting distinguishable presets needs a perceptually weighted parameter metric; the residual-difference measure used here is a starting point.
\section{CONCLUSION}
We modelled an engineer's style as a marginal distribution over effect-chain parameters and showed that personalisation then follows in closed form, a label-shift correction combining a generic content model with an engineer prior, needing no tuned weights and recovering the generic recommendation when the engineer prior equals the population. A 14-listener forced-choice test found evidence that the combined recommendations were perceptually distinguishable and that the observed effects were consistent with the predicted direction of control. The result is a proof of concept rather than a deployable system: the content signal is real but weak, and the engineer priors are synthetic rather than drawn from real console recalls.
\section{Acknowledgements}
G. Fazekas was supported by the Leverhulme Trust and the Royal Academy of Engineering under the RAEng/Leverhulme Trust Research Fellowships scheme. S. S. Vanka is funded by UK Research and Innovation [grant number EP/S022694/1] under the AI and Music Centre for Doctoral Training (AIM-CDT) at the Centre for Digital Music, Queen Mary University of London, London, UK. This work was conducted during Ms Vanka's summer internship at Yamaha RDD during the summer of 2026.
\section{Compliance with Ethical Standards}
The listening study was approved through Yamaha's internal review process. Participants were Yamaha employees aged 18 or over (audio developers, engineers, researchers and musicians) who were not involved in this project and were recruited through internal communication. Participation was voluntary and had no bearing on employment. All gave informed consent, could withdraw at any time and were not paid; responses were stored anonymously and no personal audio or other personal data was recorded. Stimuli were rendered from Mixing Secrets multitracks, used for research purposes only, and the content and population models use MedleyDB and DiffVox under its published terms. The Yamaha console recalls are proprietary and were analysed only in aggregate, with engineers anonymised; they are not released, and the style priors used in all experiments are synthetic, though grounded in directions observed in the real recalls.
\bibliographystyle{ieeetr}
\bibliography{strings,refs}
\end{document}